\documentclass[9pt]{article}

\usepackage{tikz} 
\usepackage{amsmath}

\newcommand{\rmd}{\ensuremath{\mathrm{d}}}

\title{Gravitomagnetic forces and quadrupole gravitational radiation from special relativity}

\begin{document}

\author {P.~Christillin\\ \
Dipartimento di Fisica, \\
Universit\`a di Pisa\\
I.N.F.N. Sezione di Pisa\\ \\
and\\ \\
L.~Barattini\\
Universit\`a di Pisa}

\maketitle

\begin {abstract}

The mere principle of relativity  and Lorentz transformations for the mass current predict, in close analogy to
electromagnetism, the existence of gravitomagnetic fields. With the reasonable assumption of the non existence 
of a gravitomagnetic mass, a parameter free set of equations  for  ``effective''  vector gravitoelectromagnetism
results.  As a consequence gravity propagates 
at the speed of light in a flat  Minkowski space and quadrupole gravitational radiation, consistent with energy 
balance,  is predicted with the same value of GR.
\end {abstract}

\section{Introduction}

Recently an interesting line of research has been devoted to investigate the predictions of  a set of effective vector equations \cite{veto}, derived from the basic equations of general relativity (GR) in the weak field low velocity regime.

These represent the so called gravitoelectromagnetism (GEM) formulation
\cite {lano,  bed, kol, mas, ago}.

The purpose of the present paper, which deals with the same subject, is in a sense opposite.
We are going to show that an analogous set of "Heaviside" \cite{heavi} vector equations can be derived to $O(v^2/c^2)$ from special relativity (SR) in a parameter free way.

The essential feature will be shown to be played by the relativistic mass current density which differs from the e.m. one by a factor of 2, whose role in the correct  treatment of the Lorentz force is paramount. 

The problem of the gravitational radiation, not considered by the quoted authors, will be addressed in addition. From the previous observation about the form of the mass current it will be shown that " mutatis mutandis " ( $ G\to 1/(4 \pi \epsilon_{0})$) the gravitational formula can be obtained from the e.m. one in elementary terms ($2^2=4$). 

Indeed it has long been noticed  the striking similarity between the expression for the electromagnetic quadrupole radiation
\begin{equation}
W_{el} =   \frac{1}{4\pi \epsilon_{0}} \frac{1}{180 c^5}  \left(\frac{\rmd^3 Q}{\rmd t^3}\right)^2
\end{equation}

and  the gravitational one
\begin{equation}\label{R_{S}}
W_{G} =\frac{G}{45 c^5}  \left(\frac{\rmd^3 Q}{\rmd t^3}\right)^2
\end{equation}

 Q representing respectively  the electric and gravitational quadrupole (connected to the previous one by the obvious replacement $ m \to q$).

This quantity is generally considered to be a crucial test of Einstein's General Relativity (GR) (\cite{Einstein}).
As well known, so far, only an indirect evidence of such an effect has come from the observation of binary 
pulsars (PSR B1913+16 \cite{Hul}  and PSR J0737-3039A/B \cite {kra}).
This has  strengthened the general belief of space-time curved by matter and of gravitational waves as 
"ripples which propagate in space time''  

On the contrary we will show that such an effect just comes from special relativity and that its {\it parameter free prediction } leads to an alternative simpler interpretation of physical reality.

\section{Gravitomagnetic vs. magnetic fields}

Our starting point will be the Newton attractive force of two masses $M$ and $m$ at rest (as confirmed by 
Cavendish experiments). All speculations about the scalar, vector, or tensor origin of this interaction will be for 
the moment left aside. This interaction is a matter of fact, irrespective of its more or less satisfactory implementation 
in a Lagrangian language, which is neither a necessary viaticum for truth nor a must for the description of physical reality. 

That the magnetic field comes from Special Relativity (SR) transformations of the electric field is well known. Let us 
briefly recall the pedagogical way this is presented by Feynman \cite{fey}. A charged particle $q$ is at rest at a 
distance $r$ from an infinitely long current carrying wire.  The wire is uncharged so that its positive and negative 
linear charge densities are equal
\begin{equation}\label{lam}
\lambda_{+} = \lambda_{-} = \lambda_{0}
\end{equation}
If we now let the charge $q$ move with velocity $v$ (which of course must not be equal to the drift velocity $v_{d}$ of 
the  charges of the wire, although in this case the given example requires only elementary calculations of 
immediate physical interpretation), {\it invoking the principle of relativity one predicts just from electrostatic a 
new force}  since the Lorentz transformed
\begin{equation}
\lambda'_{+} \neq  \lambda'_{-}
\end{equation}
The term coming from $\lambda'_{+}-\lambda'_{-} \simeq \lambda_{0} \; v^2 /c^2 $ can be easily interpreted as giving rise to a force 
due to the magnetic field $\mathbf{B}$ whose ``electric nature'' , in terms of the current i,  is transparent since $B = 1/(4\pi \epsilon_{0} c^2) 
2\lambda_0 v/r  = \mu_{0}/4 \pi ( 2 i/r )$.

In order to extend these elementary considerations to gravity, one can rephrase the previous case in terms of 
{\it repulsive} charges of the same sign, by substituting the positive  charges of the wire with a fictitious wire of 
negative charge density, parallel to the the first one, at the same distance from $q$ but at the opposite side.

Therefore in the gravitational case the same situation will be realized by considering a mass $m$, initially at rest, 
with two coplanar  parallel wires at the same distance $r$ and at opposite sites,  one at rest ($2$) with mass density $\lambda_{2}$ and the other ($1$) 
moving with velocity $v$,  with the same mass (or energy) density $\lambda_{1}$
\begin{equation}
\lambda_{1} =  \Delta m_{1}/\Delta  l_{1} =  \lambda_{2} =  \Delta m_{2}/\Delta  l_{2} = \lambda
\end{equation}
Hence no net force on $m$.
 
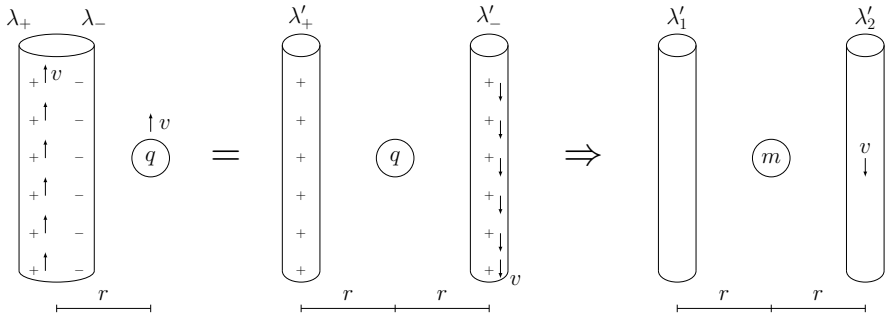
\begin{figure}[h]
\begin{center}
\scalebox{0.5}{
\begin{tikzpicture}[x=1.0cm, y=1.0cm]
\begin{scope}[>=latex]
\draw (0,0)--(0,6);
\draw (2,0)--(2,6);
\draw (0,0) arc (180:360: 1 and 0.3);
\draw (1,6) ellipse (1 and 0.3);

\draw (1,-1)--(3.5,-1);
\draw (1,-1.1)--(1,-0.9);
\draw (3.5,-1.1)--(3.5,-0.9);

\draw (2.25,-0.9) node[anchor=south] {\LARGE $r$};

\draw (3.5, 3) circle (0.5);
\draw (3.5,3) node {\LARGE $q$};
\draw[->] (3.5,3.7)--(3.5,4.2);
\draw (3.6,3.9) node[anchor=west] {\LARGE $v$};

\draw (0.4,0.0) node {$+$};
\draw (1.6,0.0) node {$-$};
\draw (0.4,1.0) node {$+$};
\draw (1.6,1.0) node {$-$};
\draw (0.4,2.0) node {$+$};
\draw (1.6,2.0) node {$-$};
\draw (0.4,3.0) node {$+$};
\draw (1.6,3.0) node {$-$};
\draw (0.4,4.0) node {$+$};
\draw (1.6,4.0) node {$-$};
\draw (0.4,5.0) node {$+$};
\draw (1.6,5.0) node {$-$};

\draw[->] (0.7,0.0)--(0.7, 0.5);
\draw (1.0,5.2) node {\LARGE $v$};
\draw[->] (0.7,1.0)--(0.7, 1.5);
\draw[->] (0.7,2.0)--(0.7, 2.5);
\draw[->] (0.7,3.0)--(0.7, 3.5);
\draw[->] (0.7,4.0)--(0.7, 4.5);
\draw[->] (0.7,5.0)--(0.7, 5.5);

\draw(0,6.3) node[anchor=south] {\LARGE $\lambda_+$};
\draw(2,6.3) node[anchor=south] {\LARGE $\lambda_-$};

\draw (5.5,3.) node {\scalebox{3}{=}};

\draw (7,0)--(7,6);
\draw (8,0)--(8,6);
\draw (7.0,0) arc (180:360: 0.5 and 0.3);
\draw (7.5,6) ellipse (0.5 and 0.3);

\draw (7.5,0.0) node {$+$};
\draw (7.5,1.0) node {$+$};
\draw (7.5,2.0) node {$+$};
\draw (7.5,3.0) node {$+$};
\draw (7.5,4.0) node {$+$};
\draw (7.5,5.0) node {$+$};

\draw(7.5, 6.3) node[anchor=south] {\LARGE $\lambda_+'$};

\draw (10, 3) circle (0.5);
\draw (10,3) node {\LARGE $q$};

\draw (12,0)--(12,6);
\draw (13,0)--(13,6);
\draw (12.0,0) arc (180:360: 0.5 and 0.3);
\draw (12.5,6) ellipse (0.5 and 0.3);

\draw (12.5,0.0) node {$+$};
\draw (12.5,1.0) node {$+$};
\draw (12.5,2.0) node {$+$};
\draw (12.5,3.0) node {$+$};
\draw (12.5,4.0) node {$+$};
\draw (12.5,5.0) node {$+$};

\draw(12.5, 6.3) node[anchor=south] {\LARGE $\lambda_-'$};

\draw (8.75,-0.9) node[anchor=south] {\LARGE $r$};
\draw (11.25,-0.9) node[anchor=south] {\LARGE $r$};
\draw (7.5,-1)--(10,-1)--(12.5,-1.);
\draw (7.5,-1.1)--(7.5,-0.9);
\draw (10,-1.1)--(10,-0.9);
\draw (12.5,-1.1)--(12.5,-0.9);

\draw[->] (12.8,0.3)--(12.8,-0.2);
\draw[->] (12.8,1)--(12.8,0.5);
\draw[->] (12.8,2)--(12.8,1.5);
\draw[->] (12.8,3)--(12.8,2.5);
\draw[->] (12.8,4)--(12.8,3.5);
\draw[->] (12.8,5)--(12.8,4.5);

\draw (13.2,-0.25) node {\LARGE $v$};

\draw (15,3.) node {\scalebox{3}{$\Rightarrow$}};

\draw (17,0)--(17,6);
\draw (18,0)--(18,6);
\draw (17.0,0) arc (180:360: 0.5 and 0.3);
\draw (17.5,6) ellipse (0.5 and 0.3);

\draw (20, 3) circle (0.5);
\draw (20,3) node {\LARGE $m$};

\draw (22,0)--(22,6);
\draw (23,0)--(23,6);
\draw (22.0,0) arc (180:360: 0.5 and 0.3);
\draw (22.5,6) ellipse (0.5 and 0.3);

\draw (18.75,-0.9) node[anchor=south] {\LARGE $r$};
\draw (21.25,-0.9) node[anchor=south] {\LARGE $r$};
\draw (17.5,-1)--(20,-1)--(22.5,-1.);
\draw (17.5,-1.1)--(17.5,-0.9);
\draw (20,-1.1)--(20,-0.9);
\draw (22.5,-1.1)--(22.5,-0.9);

\draw(17.5, 6.3) node[anchor=south] {\LARGE $\lambda_1'$};
\draw(22.5, 6.3) node[anchor=south] {\LARGE $\lambda_2'$};

\draw[->] (22.5, 3)--(22.5, 2.5);
\draw (22.5, 3) node[anchor=south] {\LARGE $v$};
\end{scope}
\end{tikzpicture}
}
\end{center}
\caption{Electromagnetic inspired derivation of gravitomagnetic effects. The neutral straight wire situation, due to 
opposite charges, is realized in gravitation with a proper placement of another mass wire. The principle of relativity 
then predicts for the moving mass $m$ an additional velocity dependent ``magnetic'' force. Relativistic gravitational 
forces can therefore be also repulsive }
\label{paragone}
\end{figure}

Let us now impart the same velocity $v$ to $m$. In its rest frame $S'$ wire $1$ will be at rest
with $\lambda'_{1} \neq  \lambda_{1}$, and $2$ will be moving with velocity $-v$, hence with
$\lambda'_{2} \neq  \lambda_{2} $  (see Fig.\ref{paragone}).
   
Now {\it the obvious but fundamental difference with respect to the e.m. case is that not only lengths shrink and 
stretch, but that also masses vary according to special relativity. Therefore the effect to first order in 
$v^2/c^2$ will be a factor of $2$ with respect to the e.m. case !} This is easy to understand since, just because of dimensional
considerations, $m$ and $l$ behave the opposite way with respect to relativistic effects.

A probably not entirely superfluous comment may be that SR would predict no gravitomagnetic force if the effects 
would subtract (1-1 = 0) instead of adding (1+1=2).
 
From
\begin{equation}
\lambda'_{1} =  \lambda_{1}/\gamma^2(v) =  \lambda (1-v^2/c^2)
\end{equation}
and 
\begin{equation}
\lambda'_{2} =  \lambda_{2}\gamma^2(v) \simeq  \lambda (1+v^2/c^2)
\end{equation}
it is then immediate to get an ``effective'' mass density difference $m$ (which is clearly not a four vector because of one  factor of $\gamma$ coming from the mass and the other one from the length ) 
\begin{equation}
\Delta \lambda' \simeq - 2 \lambda \: \gamma^2 (v^2/c^2)  
\end{equation}
which determines 
\begin{equation}\label{g'}
g' \simeq \gamma^2 \frac{4 G}{c^2}\frac{\lambda v^2}{r}  
\end{equation}
from which  one can define, paralleling the e.m. case,
\begin{equation}\label{factor2}
\mathbf{h} = \gamma^2 \frac{2 G}{c^2}  \frac{\mathbf{r}\times\mathbf{j}}{r^2}
\end{equation}
in terms of the ordinary definition of current density $\mathbf{j}= \rho \mathbf{v}= \frac{\lambda}{S}\mathbf{v}$ 
($S$ standing for the area of the wire).

The transformation properties of $\mathbf{g}$ are therefore
\begin{equation}
\mathbf{g'} = \gamma^2 \: \mathbf{v} \times \mathbf{h}
\end{equation}

Let us anticipate that the current 
\begin{equation}
\mathbf{j^*}   = 2\mathbf{j} 
\end{equation}
is the one which satisfies the continuity equation 
\begin{equation}\label{cont}
\nabla \cdot \mathbf{j^*} = -\frac{\partial \rho}{\partial t}
\end{equation}
Loosely speaking, since $\mathbf{j^*}$ and not $\mathbf{j}$ is the quantity which comes from the transformation of $\rho$, 
it is the former which combines with the latter to form, at this $(O(v^2/c^2))$ level, the required four vector in the 
continuity equation. 
In other words the mass current is \emph{not} obtained from the electric one by the simple minded replacement 
$q \to m$! If one is rightly puzzled by this result which seems to clash with the usual current continuity equation,  
let us reemphasize that this is a $v^2/c^2$ expansion of a relativistic transformation to get a $g'$  which is then 
reinterpreted as a current induced magnetic term. Thus, in a sense, this is not the standard $\mathbf{j} = m \mathbf{v}$ 
one usually deals with.

Along the same lines the same result is recovered for the infinite plane in terms of the surface mass
density $\sigma$  thus confirming in general the existence of a gravitomagnetic field $\mathbf{h}$.
 
The consideration of self energy effects, proven to be essential to provide in simple terms for the so called 
``crucial tests of GR'' \cite{christI} does not alter our conclusions.  Indeed if one uses the correct expression
\begin{equation}
g = 2 G \frac{\lambda}{r} (1- G \lambda/c^2)
\end{equation}
it is easily seen that the term in round brackets ( which disposes of space curvature ) yields terms which might 
be thought of as generating gravitomagnetic effects of higher order in $v^2/c^2$. Its contribution to $g'$ of 
Eq. (\ref{g'}) is therefore negligible. This should not come as a total surprise also in the traditional formulation 
of gravity, since vector gravity might represent the most relevant part of the energy momentum tensor. Indeed one can 
associate our ``vectors'' to the $O(1/c^2)$ components of the matter energy momentum tensor $T_{0 j}$, whereas the 
``curvature terms'' given by $T_{i j}$ are $O(1/c^4)$.

\section{The vector equations}
 
We can thus define the gravitomagnetic force which adds to the Newtonian one as
\begin{equation}
\mathbf{F}= m (\mathbf{g} + \; \mathbf{v}\times\mathbf{h})
\end{equation}
where $m$ is the relativistic mass. 

{\it We then see that, as a relativistic effect, gravitation may become repulsive.}

Thus a post Newtonian formulation of gravitation has necessarily to embody a short distance repulsion from self 
energy effects (which modifies Newton's law) and  velocity dependent possibly repulsive terms, both effects, 
somewhat at variance with the standard picture, coming from elementary considerations. 

Of course one might have defined the magnetic field without the factor of $2$,  which would then enter the 
gravitomagnetic Lorentz force (in other words it is only the combination of the expression of the gravitomagnetic field $\bf h$ and the Lorentz force which determines the dynamics). But this in turn is known (at least in e.m. and the analogy seems plausible)
to determine the Faraday induction law  which reads with our choice
\begin{equation}\label{amp}
\nabla \times \mathbf{g} = - \frac{\partial\mathbf{h}}{\partial t}
\end{equation}
and which would correspondingly change. Therefore one way or another the gravitomagnetic equations differ 
from the corresponding Maxwell ones by the factor of $2$ required by special relativity, determining at the same 
time the Lorentz force. 

For the above mentioned reasons we prefer to make the role of the modified current explicit. Thus
\begin{equation}
\nabla \times \mathbf{h} = - \frac{4 \pi G}{c^2}\,  2\,\mathbf{j}
\end{equation}
The analogy with the e.m. case, apart from the fundamental factor of $2$, is plain.  We may also parenthetically 
remark that this parameter free prediction gives a physical justification of Heaviside's Ansatz \cite{heavi} which was made,
needless to recall, prior to SR.

It is also clear that this equation holds true (like Ampere's law) only in stationary conditions. In the time 
dependent case  a \emph{gravitational displacement current} has to be introduced in order to satisfy current conservation.
The former equation will thus read 
\begin{equation}\label{spo}
\nabla \times\mathbf{h} = -\frac{4 \pi G}{c^2}\, 2\, \mathbf{j} + \frac{1}{c^2}\frac{\partial \mathbf{g}}{\partial t}
\end{equation}
where $\mathbf{g}$ represents the ``ordinary'' Newtonian field obeying
\begin{equation}\label{new}
\nabla\cdot\mathbf{g} = - 4 \pi  G  \rho
\end{equation}

These two equations are then implemented by the \emph{assumption} (a fortiori even more reasonable than 
in electromagnetism) of the non existence of a gravitomagnetic charge
\begin{equation}\label{car}
\nabla \cdot\mathbf{h} = 0
\end{equation}
which implies the existence of a gravito vector potential $\mathbf{b}$
\begin{equation}\label{pot}
\mathbf{h} =\nabla \times\mathbf{b}
\end{equation}

The implementation,  so to say, of SR in Heaviside vector formulation of gravitation \cite{heavi},  thus yields a parameter free set of equations for the weak field, low velocity gravitational case.

All discussions about the
necessity of ruling out a vector formulation of GR ( although the
present one is manifestly an approximate, of the same $O(1/c^4)$ of GR solutions),  because it
would be repulsive, seem therefore idle.

Since
\begin{equation}
\mathbf{g} =  - \nabla \phi  -\frac{\partial\mathbf{b}}{\partial t}
\end{equation}
it is immediate to get for the potentials  from the previous equations 
$(\phi, \mathbf{b})$ in the Coulomb-Newton gauge $\nabla\cdot\mathbf{b}=0$
\begin{equation}
\nabla ^2 \phi   =  4 \pi G \rho
\end{equation}
 i.e. \emph{an instantaneous gravitopotential, which thus explains the success of Newton's formulation, and 
a transverse vector potential} 
\begin{equation}
\nabla ^2\mathbf{b}-\frac{1}{c^2}\frac{\partial^2 \mathbf{b}}{\partial t^2} =  -\frac{4 \pi G}{c^2}\, 2\, \mathbf{j}
\end{equation}

The particle momentum $\mathbf{p}$ becomes
\begin{equation}
\mathbf{p} = m (\mathbf{v} + \mathbf{b})
\end{equation}
in close analogy with the e.m. case. The body $m$ gets  an extra contribution to its momentum from
the \emph{currents} through $\mathbf{b}$, in addition to the Newtonian one. 

However in our case
\begin{equation}
\mathbf{b}  =  \frac{G}{c^2} \int \frac{2 \mathbf{j}}{|r-r'|_{t'}} \rmd V'
\end{equation}

It is immediate to see, as a consequence of the previous set of equations,  that the propagation velocity of gravity
is given by
\begin{equation}
\nabla ^2-\frac{1}{c^2}\frac{\partial^2}{\partial t^2} =  0
\end{equation}

\section{Energy balance and radiation}

The continuity equation  reads
\begin{eqnarray}
& \frac{4\pi G}{c^2} \mathbf{j^*}\cdot \mathbf{g} &=  - (\nabla\times\mathbf{h})\cdot\mathbf{g} +  
\frac{1}{c^2} \mathbf{g}\cdot \frac{\partial \mathbf{g}}{\partial t}= \\
& &=- \nabla\cdot (\mathbf{h}\times\mathbf{g}) + \frac{1}{2c^2}\frac{\partial (\mathbf{g}^2+ c^2 \mathbf{h}^2)}{\partial t}
\end{eqnarray}
and the energy density is given by
\begin{equation}
U = -\frac{1}{4\pi G}\frac{\mathbf{g}^2 + c^2 \mathbf{h}^2 }{2}
\end{equation}
where the gravitoelectric energy density agrees with the
corresponding static expression. 

The energy flux $\mathbf{H}$ (duely dubbed after Heaviside) is
\begin{equation}
\mathbf{H} = \frac{c^2}{4\pi G} \: \mathbf{g}\times\mathbf{h}
\end{equation}
Therefore the energy balance comes
out right.
Notice also, as a consequence of the former wave equations, that in vacuum 
\begin{equation}
h  = g/ c   
\end{equation}
Therefore in vacuum
\begin{equation}
U  = -\frac{1}{4\pi G} g^2 
\end{equation}
and 
\begin{equation}
|H| =\frac{c}{4\pi G}  g^2
\end{equation}
so that
\begin{equation}
|H| = c \: U 
\end{equation}
i.e. the usual relation one has between flux and energy density for electromagnetism.

It is easy to see that energy would be conserved \emph{ even} with the coefficient $2$ in front of the 
displacement current.  This would however imply a propagation velocity $c^*=c/\sqrt {2} $, thus violating 
causality and affecting radiation where \emph{also the  power depends on the propagation velocity}.  
 
Thus we straightforwardly  predict a quadrupole radiation
\begin{equation}
W_{G} = W_{el}\,\left(\frac{1}{4\pi \epsilon_{0}}\to G,\  q \to 2 m \right) 
\end{equation}

\emph{In conclusion the factor of $4$ in the quadrupole gravitational radiation, traveling with speed $c$ in 
a flat Minkowski space, can be simply derived by these elementary considerations  based entirely on 
special relativity.}

 As a consequence also in all other multipoles the same replacement takes the place of the naive $ q \to m$.

\section{Comments and conclusions. }

In the present paper the unavoidable contribution of special relativity to gravitation has been calculated.

It has been shown that SR plays a paramount role also in what has been considered so far to be a distinctive 
feature of gravitation i.e. radiation. This should not come as a total surprise also in the traditional 
formulation of gravity, since vector gravity might represent the most relevant part of the energy momentum tensor. 
Indeed one can associate our ``vectors'' to the $O(1/c^2)$  components of the matter energy momentum tensor 
$T_{0 j}$, wheres the ``curvature terms'' given by $T_{i j}$ are $O(1/c^4)$.

Some comments as regards the results obtained and their relation to related works are in order.
 
Indeed in the literature different versions of the so called GEM (gravitoelectromagnetism) equations are available 
\cite {lano,  bed, kol, mas, ago}. All of them are said to be obtained from the basic equation of GR ``assuming a 
weak gravitational field or reasonably flat spacetime'' . However the coefficient of the current in the 
4th equation varies from $1$ to $2$ to $4$.
None of them in addition  deals with the present issue i.e. the problem of the propagation 
velocity and the consistent evaluation of the ensuing gravitational radiation. All of them are incomplete 
and/or wrong (remember the previous comments about the the necessarily constrained form of the Lorentz force and of 
the induction law). 

In the case the current coefficient is $1$, which corresponds to the naif Heaviside case, the propagation velocity is 
clearly $c$, but the quadrupole radiation is $1/4$ of the  one predicted by GR. Therefore the reduction violates both 
GR and SR, since the relativistic mass variation implies the coefficient to be $2$!

If the coefficient is $2$, as  has been derived in the present work from SR, and the same factor multiplies the 
displacement current the propagation velocity would be $c/\sqrt{2}$ which manifestly violates causality. 
In addition one would predict a bigger quadrupole radiation by the same amount $\sqrt{2}$.

If the coefficient is $4$ (the extra factor of $2$ being claimed to come from spacetime distortion) the propagation 
velocity is $c/2$ and quadrupole radiation would be $16$ times bigger than predicted by GR!

In conclusion both the expressions for the periastron precession 
\begin{equation}
\frac{\Delta \phi}{\phi} = 3 G \frac{M_{1} + M_{2}}{c^2 a (1-e^2)}
\end{equation}
and 
\begin{equation}
W_{Q}  =  32 G^4  \frac{(M_{1} M_{2})^2(M_{1} + M_{2}}{5 c^5a^5 (1-e^2)^{7/2}} \times \left(1+\frac{73}{24}e^2 
+ \frac{37}{96} e^4 \right)
\end{equation}
based on the quoted expression for quadrupole radiation, basis for the PN description of the periastron advance and 
period $T$ decrease of binary systems (Shapiro effects deriving trivially from space-time ``curvature'') are easily 
derived from elementary considerations.   The first one from self energy  which affects Newton's law and 
angular momentum  \cite{christI}.
 
In this connection it is instructive to recall the beautiful example by Feynman of the bugs on a hot plate. 
If they unaware of the temperature gradient they will conclude that that they live in a curved (non euclidean) space, 
which will on the contrary appear flat to us who know of the temperature effects.  In the same way if we ignore 
self energy effects, we will attribute to space time curvature our ignorance. The two physical descriptions of our 
world are of course equivalent, apart from physical foundedness, simplicity and from the problem of energy and 
momentum conservation in a curved space time. 
 
Self energy effects have also been seen to play a paramount role in an alternative consistent cosmological 
description \cite{christII}.

In conclusion the line of research to derive GR from SR started by Schiff \cite{schi} and pursued in \cite{christI} for 
Mercury precession has been continued in the present work. 

The factor of $2$ in the current and therefore in the quadrupole is thus seen to come from an (almost) elementary 
twofold consequence of Lorentz transformations and not \emph{necessarily} from a hypothetical spin $2$ nature of 
the graviton.

The prediction  of the geodetic precession and of  frame dragging as a particular,  i.e.  stationary,  case of the present equations is being considered elsewhere. 

\vspace{0.5cm} 

ACKNOWLEDGMENTS
  
It is a pleasure (P.C.) to thank G. Morchio, P.G. Prada Moroni, G. Cicogna,  M. Lucchesi and C.Bonati for their interest 
and help in this work.

\end{document}


\title{Adddendum 1 to "Gravitomagnetic forces and quadrupole .. " 
\\
The gravitomagnetic equations and the  Heaviside Lorentz  force  }

\author {P. Christillin\\ \
Dipartimento di Fisica, \\
Universit\`a di Pisa\\
I.N.F.N. Sezione di Pisa\\ \\
and\\ \\
L. Barattini\\
Universit\`a di Pisa}

\maketitle

\begin{abstract}
Different sets of  Maxwell-like equations of gravity are given in literature under the name of  gravitomagnetism for the low velocity weak field limit of GR . We are going to show that these versions are not consistent with fundamental principles, namely interaction velocity c and  matter current conservation, which we use to derive the correct ones   accounting for all the experimental results.
\end{abstract}

\

  e-mail christ@df.unipi.it

\section{Introduction}

 In 1893 Heaviside proposed an analogy between electromagnetic and gravitational fields, as both described by formally equal Maxwell equations \cite{heavi}.
 
Being prior to special relativity (SR), the two fields (gravitoelectric henceforth denoted by $\bf g$ and gravitomagnetic by $\bf h$ ) were thoght to be independent and the propagation velocity  also appeared as a free parameter. 
 
Using special relativity (SR)  an "improved" version of the Heaviside equations, of use where non linear effects do not come into play (light deflection and the periastron precession),  has been  recently derived from us \cite{christbara}. 
 
 The aim of the present paper is to obtain the same result from more general arguments, against analogous sets of equations derived in the weak field low velocity case from GR \cite{lano} \cite{bed} \cite{kol} \cite {mas}  \cite{ago}.

\

\section{The vector equations}

Since, as mentioned,  the  Heaviside-inspired equations  differ from different versions of analogous equations derived, in the same conditions, from General Relativity, it is  mandatory to justify them.

The vector equations for gravitation are the following : 
\begin{equation}\label{new}
\nabla\cdot\mathbf{g} = - 4 \pi  G  \rho
\end{equation}
\begin{equation}\label{car}
\nabla\cdot\mathbf{h} = 0
\end{equation}
where in the first $\mathbf{g}$ represents the "ordinary'' Newtonian field, while  the second for the gravitomagnetic 
field $\mathbf{h}$ is based on the  \emph{assumption} of the 
non existence of a gravitomagnetic charge.
 
These two are accompanied by the time dependent ones :
\begin{equation}\label{amp}
\nabla \times \mathbf{g} = -\frac{\partial\mathbf{h}}{\partial t}
\end{equation}
\begin{equation}\label{spo}
\nabla \times\mathbf{ h} = -\frac{4\pi G}{c^2}\,2\, \mathbf{j} + \frac{1}{c^2}\frac{\partial \mathbf{g}}{\partial t}
\end{equation}
where
\begin{equation}\label{cu}
\bf{ j} = m \bf{v}
\end{equation}
Eq.(4)  differs from the corresponding Maxwell one by the factor of $2$ in front of the ordinary 
mass current density  $\mathbf{j}$.

Thus a post Newtonian formulation of gravitation has necessarily to embody a short range repulsion (which modifies Newton's law) from self energy 
effects of higher order in $1/c^2$ not considered here \cite{christI}  and velocity dependent, possibly repulsive terms, both effects, somewhat at 
variance with the standard picture.

{\it  Our aim will be simply to show how, under the same conditions, the previous equations can be justified  without having to resort to a specific theory }(in particular GR).

\

  \
  
  i) In Eq.4) $\nabla \times\mathbf{ h}$ is determined by $ 2\mathbf{j}$.
  
  \
  
  That the mass current, unlike the electric current which is a 4-vector, is the $T_{oi}$ component of the energy momentum tensor $T_{\mu\nu}$ is well known.

 An undisputed requirement is that this quantity has to be conserved.  That this does not necessarily correspond to the naive constraint 
  \begin{equation}\label{new}
k_{\mu}T_{\mu\nu} = 0 
\end{equation}
is clear, since the ordinary derivative differs from the covariant one.

  This does not come as a surprise from the QED and QCD cases, and can be considered "granted" here, even without an explicit theoretical formulation, because of the non linearity of gravitation (non abelian theory because of the self coupling of the graviton).  Moreover the information one would get from the previous equation by separating the time and  spacial components of the free $\nu$ index are not covariant.

Indeed an explicit example of the correctness of the previous statement is given by Feynman \cite{fey} when considering Compton the scattering of gravitons. 

Notice that the previous condition is used in GR for the calculation of the quadrupole gravitational radiation, which will be commented upon later on, only as a result of the equations of motion. This is what we try to avoid.

Thus we seek which information can be derived from the condition
  \begin{equation}\label{new}
k_{\mu} k_{\nu}T_{\mu\nu} = 0
\end{equation}
which overcomes some of the previous problems and which appears to be unquestionable. Thus 
 \begin{equation}\label{new}
k_{\mu} k_{\nu}T_{\mu\nu} = 0 =   \omega^2 T_{00} +   \omega T_{0i}k_{i}  +  \omega k_{i} T_{i0} +  k_{i}k_{j} T_{ij} 
\end{equation}
 Given the fact that $T_{ij}$ is one higher order in $1/c^2$  than $T_{0j}$ and the symmetry $T_{0i}=T_{i0}$, we thus have 

  \begin{equation}\label{new}
  \omega T_{00} +  2   T_{0i}k_{i}  \simeq 0 
\end{equation}

or 
\begin{equation}\label{cont}
\frac{\partial \rho}{\partial t} = -\nabla \cdot 2\mathbf{j} = -\nabla \cdot 2 m \mathbf{v}
\end{equation}

Thus mass current conservation gets from first principles  the "strange" coefficient 2, explicitly obtained in gravitomagnetism from SR transformations of the mass density \cite{christbara}, paralleling the analogous e.m. current case, where on the contrary 
or 
\begin{equation}\label{cont}
\frac{\partial \rho_{em}}{\partial t} =  = -\nabla \cdot q \mathbf{v}
\end{equation}

This represents the explanation and the implementation of the so called  "spin two nature of the graviton".

\

ii) The displacement current, given the unquestioned validity of Eq.1), must appear in Eq.4) in the given form.

Only in this way, by taking  the divergence of both sides,  can it comply with current conservation . 

\

iii) The displacement current in Eq.4) and the the gravitomagnetic field in the Heaviside-Faraday Eq.3) {\bf must} appear in the given form. Indeed only so the free field equations for ${\bf g}$
\begin{equation}\label{new}
 \frac{1}{c^2}\frac{\partial^2 \mathbf{g}}{\partial t^2} - \nabla\mathbf{g} = 0
\end{equation}
  and ${\bf h}= \bf{g}/c$ obtain. 
  
  {\it This implies, as due, a propagation velocity $c$ for gravitation.}
  
\

iv) Eq.3) determines the   Heaviside-Lorentz gravitomagnetic force
\begin{equation}\label{lor}
\mathbf{F} =   m (\mathbf{g} + \mathbf{v} \times \mathbf{h})
\end{equation}
where $m$ is the relativistic mass.

 The induction law 
in its integral formulation, for the case of constant $\mathbf{h}$ and a varying circuit is in agreement with the 
Lorentz force \emph{only} in the present form. The same holds true for a constant area and varying flux so that finally  the differential form is recovered \cite{berk}.

This represents therefore a double confirmation of the present formulas.

\

v) The radiation fields being determined by $ 2\mathbf{j}$, 
the  energy flux $\mathbf{H}$ 
\begin{equation}
\mathbf{H} = \frac{c^2}{4\pi G} \: \mathbf{g}\times\mathbf{h}
\end{equation}
is 4 times bigger than it would be, had one used $\mathbf{j}$. Thus in comparing gravitational to electromagnetic radiation the naive replacement $ q \to m $  has to be corrected by the said factor.  Of course this applies to the corresponding multipoles in the two cases and since for gravity no dipole and magnetic dipole emission is possible,
 we straightforwardly  predict a quadrupole radiation
\begin{equation}
W_{G} = 4 \; W_{el}\,(\frac{1}{4\pi \epsilon_{0}}\to G)
\end{equation}

This rests of course also on point i). 

\

Consistency is obtained by taking the previous arguments, at pleasure, also in the reverse order. 

One can prefer to state the $\bf g$ and  $\bf h$ can enter Eq. (3) and (4)  only in the given form in order to get a propagation velocity c , then that  $\bf j$ has to enter Eq.(4)  with a factor of 2 to enforce current conservation;  these equations then determine the Heaviside-Lorentz force as given in Eq.(13).

Thus the equations  coming from the GR reductions in  \cite{lano} \cite{bed} \cite{kol} \cite {mas}  \cite{ago}, which differ from the present ones by some coefficients violate at least one of the previous requirements. 

\

The previous set of equations ( unlike Maxwell  invariant under Lorentz transformations only at the $O(v^2/c^2)$ level )  cannot be applied to objects moving around the earth which is not an inertial frame .

We have to add the Coriolis (and centrifugal) force

\begin{equation}
\mathbf{F}_{Cor}  =   2 m \, \mathbf{v} \times\mathbf{\omega}
\end{equation}

where $\omega$ refers respectively to each of the two rotations induced by the movement of the earth on the moving object.

Its cosmological explanation can be obtained by considering it to be due to the counterrotation of  the (rest of the) Universe.  Thus from the gravitomagnetic field of
a mass $m$, which represents a N.R. reproduction of the Lense-Thirring effect,  it  follows
\begin{equation}\label{bri}
\mathbf{F}_{GM}  =    m\, \mathbf{v} \times\left( \frac{2 G M}{c^2 R}\mathbf{\omega}\right) = 2 mv \frac{GM}{c^2R} 
\omega \, \mathbf{n}
\end{equation}
the suffix GM standing for gravitomagnetic, and if 
\begin{equation}\label{U}
\frac{GM_{U}}{c^2 R_{U}} =1 
\end{equation}
then
\begin{equation}
\mathbf{F}_{Cor}  =   \mathbf{F}_{GM} 
\end{equation}

Thus the gravitomagnetic "Lorentz" force for the earth can be written, as done in the literature from people who "derive" it from the GR equations  \cite{schi} \cite{mach}, as
\begin{equation}\label{lor}
\mathbf{F} =   m (\mathbf{g} + 2 \;\mathbf{v} \times \mathbf{h})
\end{equation}

Coming  to torques, again a Coriolis contribution has to be added to the  gravitomagnetic field $\mathbf{h}$ so that the equation of motion  of the satellite gyroscope (of standard angular momentum 
$\mathbf{S} = m r^2 \omega_{orb} \,\mathbf{n} =  2\mathbf{\mu}$)  then reads in the non inertial frame of the earth 
\begin{equation}\label{torque}
\frac{\mathbf{S}}{2} \times ({\mathbf{h}} +\mathbf{\omega})  = \frac{\rmd\mathbf{S}}{\rmd t} 
\end{equation}
where $\bf{\omega}$ refers to the earth rotation and revolution.

Let us stress once more that in the case of the gyroscopes of  orbiting satellites  \emph{the fact that 
forces are locally eliminated in the free falling frame  (no tide effects), does not imply the same for the moments !}.  

Thus the Gravity Probe B experiments \cite{gravity} prove that for gravity no "true" inertial frame exists !

The same happens, in principle, also for the earth : precession and nutation, although determined by tide effects, get also a (much smaller) contribution from gravitomagnetism.
\

\section{Conclusions}

In the  present paper  we have clarified the issue of  the different versions of the gravitomagnetic equations and their relation to the Heaviside-Lorentz force providing the correct ones just from elementary considerations. GR has been shown to be irrelevant for their derivation.

\


\author { \\P. Christillin\\ \
Dipartimento di Fisica, \\
Universit\`a di Pisa\\ and \\ I.N.F.N.  , Sezione di Pisa}

\maketitle

\begin {abstract}

Heuristic considerations relate gravitational radiation to the strong field parameter $\frac {GM}{c^2 r}$ , suggesting the complete evaporation of a black hole at the Planck scale.

\end {abstract}			
				
	\

	e-mail christ@df.unipi.it

\

{\bf Gravitational power}

\

It is usually said that gravitational radiation be a proof of General Relativity. 

That this is not so has been  shown in \cite {christ I}.

The striking similarity of gravitational quadrupole 
to electric quadrupole radiation
had indeed been noted. However this was regarded no more than a curiosity since no link between the two could be made.  
On the contrary in the derivation of effective vector equations valid for gravitation in the weak field limit, which is undoubtedly the case of radiation, it has been shown that   the tensor nature of the mass current plays a paramount role.  The mass current enters the  continuity equation and the corresponding  Heaviside-Lorentz equations with a factor of two with respect to the electric case. Thus 
given  the obvious ratio of the coupling constants $G/4\pi \epsilon_{0}$, only a factor of 4 results in the ratio of the gravitational to electric quadrupole radiated power.

Let us now elaborate on the expression for quadrupole power

\begin{equation}\label{R_{S}}
 W_{G} = G /(45 c^5 )  (d ^3 Q/dt^3)^2 
 \end{equation}

\
First of all the matrix element can  be got straightforwardly from dimensional considerations, since  
 the static Newtonian field has to go into the radiative one as 
 
 \begin{equation}\label{la}
g =  \frac{GM}{r^2}  \to  g_{rad} = \frac{G}{r} ( \frac{dM}{c dt} + \frac{d^2Ml}{c^2dt^2} + \frac{d^3 (Ml^2)}{c^3 dt^3} ) 
 \end{equation}
where the monopole and dipole terms are absent for  well known reasons.

The matrix element of the gravitational power has the dimensions of a power.  As a matter of fact 

\begin{equation}\label{vecpot}
 \frac{d^3 (Ml^2)}{dt^3}  \simeq  \frac{  Md^2 l }{dt^2 } \times  \frac{d l}{dt} \simeq M a \times v  =  F \times v  \simeq  \frac{GMm }{r^2} \times v
 \end{equation}
  where $l$  stands  for the dimensions of the source, $a$  and $v$ for its acceleration and velocity (we have not distinguished here between total mass M of the system and individual masses M and m in the force, neither minded  that, as vectors, F and v are orthogonal for a circular orbit. Of course this way of splitting the derivatives is not unique; traditionally it is expressed as the time derivative of the kinetic energy, which dimensionally amounts to the same thing. We preferred this formulation which privileges the role of the force.).
  
 Thus the quadrupole radiation power  is simply due to the Newtonian orbiting force times the orbiting velocity : a quite intuitive elementary interpretation.

Since, in addition to Special Relativity (SR) necessary for the derivation of the effective vector equations used to predict the previous result,  self energy is the other necessary ingredient which accounts for the non linearity of gravitation it is mandatory to ascertain its relevance even in this case. 

Self energy is represented in this case by radiated energy within the sphere of radius $r $ (the distance of the emitter  from us)  and it 
can be easily obtained as the  the radiated power multiplied by the time it takes radiation to fill up the given sphere i.e. $ E_{r}  =  W_{G}\times (\frac{ r}{ c })$ ( we are simplistically assuming a constant power) .  We thus get   a "strong field parameter "  which would correct the flat space term $g_{rad}$ by 

\begin{equation}
 g_{rad}  \to g_{rad}(1-  \frac{G M_{r}}{ c^2 r }) = g_{rad}(1-  \frac{G }{ c^5 } W_{G})
\end{equation}

   where $G/c^5 \simeq 10^{-52}  W^{-1} $.  In the presently available cases (for example for the binary system $B1913+16 $, $W_{G} \simeq 5 \times 10^{24}$ W) the effect is really tiny.
  Thus the  correction is extremely small so that forgetting self energy effects  is  absolutely legitimate. 

 But what in the general case ?
 
 It is  obvious (since the matrix element has the dimensions of a power) and well known that $c^5/G$ has  also the dimensions of a power. Probably  it is less recognized it represents {\it the Planck power $W_{P}$ }

\begin{equation}\label{R_{S}}
\frac{c^5}{G} = \frac{E_{P}}{t_{P}} = \frac{\sqrt \frac{\hbar c^5}{G}}{\sqrt \frac{\hbar G}{c^5}} = W_{P}
 \end{equation}
 
 where {\it $\hbar$ disappears} !

We thus see that  the usual weak field parameter which entered the stationary case also enters the radiative one in such a way that 

\begin{equation}
\frac{G M_{r}}{ c^2 r } = 1 \Rightarrow  \frac{G }{ c^5 } W_{G} =  \frac{W_{G}}{W_{P}} =1
\end{equation}

i.e.  its limiting strong value of 1 corresponds to the maximum possible value of the emitted power $W_{P}$. 

In other words no more power can be radiated off (all energy having already been disposed of).

This happens for $v=c$ at 

\begin{equation}\label{vecpot}
F_{N} = \frac{GMm }{r^2} = \frac {c^4}{G} = F_{P}
 \end{equation}
 i.e. when Newton's force has reached the Planck value. 

 The "true" power expression thus reads

 \begin{equation}
 W^*_{G}  =  W_{G}(1-  \frac{W_{G}}{W_{P}})^2
\end{equation}

\

{\bf Comments and conclusions}

\

 Should these speculations be taken earnestly ? 
 
 The problem of black holes is indeed a very old one.

In his fundamental article of $1939 $ Einstein  \cite {ein}  dismissed the possibility of the existence of black holes (at that time called the Schwarzschild singularity \cite {ss}) . 

This work was later commented upon by Thorne \cite {Fey} as a " regrettable paper "  ; the scientific community had decided for their existence and reneged the founder ! 

Among others \cite {bar} the matter was recently taken up by the present author who, on the contrary, stubbornly \cite {Fey} " resisted the outrageous implications of Schwarzschild's solution "  showing that self energy should swallow the mass of the  object. 
This was based on a first order treatment of the self energy.  Along this line a more accurate  self consistent  consideration of the self energy by Dillon \cite {dil} has led to the same conclusions. 

The present considerations are interesting and questionable in many respects. 

First they link gravitation with quantum mechanics through the only Planck quantities where $\hbar$ has disappeared, thus raising an intriguing question. Is gravitation, though consistent with QM, really independent of it ? 

As a matter of fact in \cite {christ IV} it had been argued that at the Planck scale one has 

\begin{equation}
E_{P} -   \frac{G M_{P}^2}{R_{P}}  =0 
\end{equation}

as it appears to happen also for our Universe (U)

\begin{equation}
E_{U} -   \frac{G M_{U}^2}{R_{U}}  =0 
\end{equation}

suggesting that we might be living in a black hole where particle creation should have occurred at zero cost, due to the black body gravitational self energy.

Here the situation might be just reversed. 

Gravitational structures which have evolved from fluctuations might in the end disappear having radiated away all of the initial energy. The latter should furnish the dark energy, only wide spread remnant of the final non existing black hole. 

 Thus what are people talking about when they claim the presence of black holes \cite {mac} ? They are simply unduly naming so clusters of missing masses necessary to account for galaxies rotations curves with a value of $GM/c^2r$ far below 1.

 Coming to technicalities  the weakest point is surely the use of an essentially  "Newtonian  law" even in extreme conditions. 
However when it comes to facts,  this is also what happens for the Schwarzschild solution.

 As a matter of fact the latter coincides with the famous SR $1/ \gamma $ factor  

\begin{equation}\label{psi}
\sqrt{1-v^2/c^2}  =\sqrt{ 1- 2 \int {\bf F} \cdot  d{\bf r} /m c^2} = \sqrt{1- \frac{2 Gm}{c^2 r}} 
\end{equation}

Possible improvements have been critically discussed in \cite{christ III}.

In conclusion, even with the preceding provisos,  the present result connecting radiated energy to the static self energy may suggest an interesting scenario. 

\

{\bf Acknowledgments}

\

I wish to thank L. Barattini, C. Bonati, G.  Cicogna and G. Morchio for fruitful discussions and a critical reading of the manuscript.